\documentclass[journal=jacsat,manuscript=reprint]{achemso}
\usepackage[version=3]{mhchem} 
\usepackage{adjustbox}
\usepackage{xcolor}
\usepackage{hyperref}
\usepackage{cleveref}
\DeclareUnicodeCharacter{2212}{-}

\title[An \textsf{achemso} demo]
 {Stability of adsorption of Mg and Na on sulfur-functionalized MXenes}

\author{G. Chaney}
\affiliation{Department of Physics, University of Maryland Baltimore County, 1000 Hilltop Cir, Baltimore, MD 21250, USA}
\author{D. Çakır}\email{deniz.cakir@und.edu}
\affiliation{Department of Physics and Astrophysics, University of North Dakota, Grand Forks, North Dakota 58202, USA}
\author{F. M. Peeters }
\affiliation{Department of Physics, University of Antwerp, Groenenborgerlaan 171, 2020 Antwerpen, Belgium}
\author{C. Ataca}\email{ataca@umbc.edu}
\affiliation{Department of Physics, University of Maryland Baltimore County, 1000 Hilltop Cir, Baltimore, MD 21250, USA}

\begin{document}

\date{\today}

\begin{abstract}
Two-dimensional materials composed of transition metal carbides and nitrides (MXenes) are poised to revolutionize energy conversion and storage. In this work, we used density functional theory (DFT) to investigate adsorption of Mg and Na adatoms on five M$_{2}$CS$_{2}$ monolayers (where M= Mo, Nb, Ti, V, Zr) for battery applications. We assessed the stability of the adatom (i.e. Na and Mg)-monolayer systems by calculating adsorption and formation energies, as well as voltages as a function of surface coverage. For instance, we found that Mo$_{2}$CS$_{2}$ cannot support a full layer of Na nor even a single Mg atom. Na and Mg  exhibit the strongest binding on Zr$_{2}$CS$_{2}$, followed by Ti$_{2}$CS$_{2}$, Nb$_{2}$CS$_{2}$ and V$_{2}$CS$_{2}$. Using the nudged elastic band method (NEB) we computed promising diffusion barriers for both dilute and nearly-full ion surface coverage cases. In the dilute ion adsorption case, a single Mg and Na atom on Ti$_{2}$CS$_{2}$ experience $\sim$0.47 eV and $\sim$0.10 eV diffusion barriers between the lowest energy sites, respectively. For a nearly full surface coverage, a Na ion moving on Ti$_{2}$CS$_{2}$ experiences a $\sim$0.33 eV energy barrier, implying a concentration dependent diffusion barrier. Our molecular dynamics results indicate that three (one) layers (layer) of Mg (Na) ion on both surfaces of Ti$_{2}$CS$_{2}$ remain stable at T=300 K. While, according to voltage calculations, Zr$_{2}$CS$_{2}$ can store Na up to three atomic layers, our MD simulations predict that the outermost layers detach from Zr$_{2}$CS$_{2}$ monolayer due to weak interaction between Na ions and the monolayer.  This suggests that MD simulations are essential to confirming the stability of an ion-electrode system - an insight that is mostly absence in previous studies.  
\end{abstract}

\section{Introduction}
 MXenes comprise a family of two-dimensional, transition metal carbides and nitrides with a general formula of M$_{n+1}$X$_{n}$T$_{n}$ ($n$ = 1, 2, or 3) where M is a transition metal, X is a C and/or N atom, and T is a surface termination group \cite{formula}. MXenes are promising candidates for energy storage applications \cite{li_storage_1, li_storage_2, li_storage_3, li_storage_4, supercapacitor,review_sun} owing to their unique properties such as the ability to store metal ions between layers, and high charge storage capacity per unit mass and volume. Importantly, they have high electronic conductivity, which is a necessary condition to realize fast charge-discharge rates for high-rate electrodes \cite{high-rate}.  

A major focus of MXene research concerns the functionalization of the highly reactive bare structure with surface terminations (traditionally O, OH, and F)\cite{sulfer}. Apart from these surface terminations, recently, the synthesis of MXenes with O, NH, S, Cl, Se, Br and Te surface terminations has been demonstrated as well. Recently, S-functionalized MXenes have been shown via first-principles calculations to have lower Li/Na ion diffusion barriers than O-functionalized MXenes\cite{sulfer,WANG2020227791}, implying that S-terminated MXenes are strong candidates as electrode materials for high-rate Li-ion and Na-ion batteries.

In addition to surface terminations, the type of intercalation ion determines an MXene's efficacy as an electrode material.  The interactions between the surface terminations and intercalation ions can be complex.  For instance, O-functionalized MXenes remain stable upon Na- and K-ion adsorption, but may shed their O terminations upon Mg-, Ca-, or Al-ion adsorption \cite{ions}.  Therefore, it is critical to computationally test many combinations of MXenes, surface terminations, and adatoms before considering experimental realization. Our previous study performed first principles calculations on the structural stability, and the electrochemical and ion dynamic properties of Li adsorbed sulfur-functionalized group 3B, 4B, 5B, and 6B transition metal (M)-based MXenes (i.e., M$_2$CS$_2$ with M= Sc, Ti, Zr, Hf, V, Nb, Ta, Cr, Mo, and W) \cite{deniz}. In this work, we considered Mg and Na ions due to double valency of the former and abundance of the latter ion. We found that an adatom's adsorption energy, charge transfer, and distance from the surface of MXene depend on the material and adatom type.  For instance, among the considered MXenes (M$_2$CS$_2$; M = Mo, Ti, V, W, Zr), Mg and Na consistently bind least strongly to Mo$_2$CS$_2$ than to any other structure.  We performed cluster expansion calculations to obtain the lowest energy adatom configurations and open circuit voltages at different concentrations, as well as nudged elastic band calculations to compute the diffusion barriers and paths of the adatoms across the materials. In addition to the energetic, we performed molecular dynamics simulations to asses the stability of the considered systems at room temperature.

\section{Computational Methods}

To determine the stability of the examined structures, we calculated adsorption and formation energies and voltage. Adsorption energy is how strongly an adatom binds to a monolayer.  The adsorption energy ($E_{ads}$) per adatom (A) is calculated using the following expression:

\begin{equation} \label{adsEnergy}
    E_{ads} = \frac{1}{n}[E_{\mathrm{M}_{2}\mathrm{C}\mathrm{S}_{2} + n\mathrm{A}} - nE_{\mathrm{A}} - E_{\mathrm{M}_{2}\mathrm{C}\mathrm{S}_{2}}]
\end{equation}
where $E_{\mathrm{M}_{2}\mathrm{C}\mathrm{S}_{2} + n\mathrm{A}}$ is the total energy of the MXene with adatoms, $E_{\mathrm{M}_{2}\mathrm{C}\mathrm{S}_{2}}$ is the total energy of the MXene, $E_{\mathrm{A}}$ is the energy of a free adatom, and $n$ is the number of adatoms in the simulation cell.  Formation energy is the tendency of an adatom to bind to the monolayer rather than detach and form a crystal with other adatoms of the same species.  It uses the energy of the adatom's bulk structure.  For our study, we calculated the formation energy as

\begin{equation} \label{formationEnergy}
    E_{f} = [E_{\mathrm{M}_{2}\mathrm{C}\mathrm{S}_{2} + n\mathrm{A}} - nE(A) - E_{\mathrm{M}_{2}\mathrm{C}\mathrm{S}_{2}}]/n
\end{equation}
where $E(A)$ is the energy per atom of the lowest energy bulk structures of Mg or Na.  Note that according to Eq. \ref{formationEnergy}, the stable systems have negative formation energies.  Finally, the open circuit voltage is an important quantity and we investigated the half cell reaction given below as

\begin{equation} \label{half-cell}
    \mathrm{A}_{x_1}\mathrm{M}_{2}\mathrm{C}\mathrm{S}_{2} + red{(x_2-x_1)}\mathrm{A} \xrightarrow{} \mathrm{A}_{x_2}\mathrm{M}_{2}\mathrm{C}\mathrm{S}_{2}
\end{equation}
where A is either the Na or Mg adatom, and $x_1$ and $x_2$ are the numbers of adsorbed adatoms (per formula unit) before and after the reaction, respectively. The average voltage is thus computed as:

\begin{equation} \label{voltage}
    \Bar{\mathrm{V}} = -\frac{E(\mathrm{A}_{x_2}\mathrm{M}_{2}\mathrm{C}\mathrm{S}_{2}) - E(\mathrm{A}_{x_1}\mathrm{M}_{2}\mathrm{C}\mathrm{S}_{2}) - (x_2-x_1)E(\mathrm{A})}{(x_2-x_1)e}
\end{equation}
where $x_2 > x_1$, $E(\mathrm{A}_{x_1}\mathrm{M}_{2}\mathrm{C}\mathrm{S}_{2})$ and $E(\mathrm{A}_{x_2}\mathrm{M}_{2}\mathrm{C}\mathrm{S}_{2})$ are the energies of the electrode before and after the reaction, and $e$ is the unit electronic charge. According to our definition,  a positive $\Bar{\mathrm{V}}$ implies energetic stability of adsorption on the MXene surface. Negative voltages indicate that the adatoms would rather form clusters with each other than bind to the monolayer. \cite{voltage} 

Calculations of the above quantities were performed under the framework of the density functional theory, using the Vienna ab initio simulation package (VASP)\cite{abinitio_1, abinitio_2, abinitio_3, abinitio_4}. A plane-wave energy cutoff was set to 500 eV. The electron- ion interactions were described by the projected augmented wave (PAW) method\cite{paw_1, paw_2}. A generalized gradient approximation (GGA) based exchange-correlation potential was considered with the Perdew-Burke-Ernzerhof (PBE) pseudopotential\cite{gga_1}. In addition, the strongly constrained and appropriately normalized (SCAN)\cite{scan} functional was used for the single atom adsorption cases, as SCAN tends to give more accurate results for lattice constants than PBE does. We took into account van der Waals (vdW) interactions within DFT+D3 formalism.\cite{dft+d3}  Spin-polarization was set for every calculation, but spin-orbit coupling was neglected. A vacuum space larger than 15 {\AA} was used to avoid spurious interaction between monolayers. $\Gamma$-centered 12$\times$12$\times$1 and 4$\times$4$\times$1 Monkhorst-Pack k-point meshes were considered for 1$\times$1$\times$1 
(primitive cell) and 4$\times$4$\times$1 supercell calculations, respectively.  The climbing image nudged elastic band (CI-NEB) method, as implemented in the VASP transition state tools, was applied to estimate the minimum energy diffusion paths.\cite{neb_1, neb_2}  The cluster expansion method was used to predict and to test the stability of various coverages, for up to three layers of adatoms\cite{avdw:atat2,avdw:atat,avdw:maps}.  From these calculations, we acquired our voltage values. Molecular dynamics simulations were performed for Ti$_{2}$CS$_{2}$ and Zr$_{2}$CS$_{2}$ for various coverages to determine the temperature stability as a function of coverage.  We employed an isothermal−isobaric (NPT) ensemble for \emph{ab initio} molecular dynamics (MD) simulations\cite{langevin1, parrinello_1980, parrinello_1981}. We only applied the constant pressure algorithm to the two lattice vectors parallel to the 2D plane, leaving the third vector unchanged during the simulation. We kept the external pressure at 0 Pa. Our ab initio MD simulations lasted 15 ps with a time step of 1 fs, and the temperature was kept constant at 300 K. We tested different friction coefficients of atomic and lattice degrees of freedom, namely 5 and 10 ps$^{-1}$.

\section{Results and Discussion}

\subsection{Adsorption on M$_2$CS$_2$ Monolayers}

\begin{figure}
\includegraphics[width=13cm]{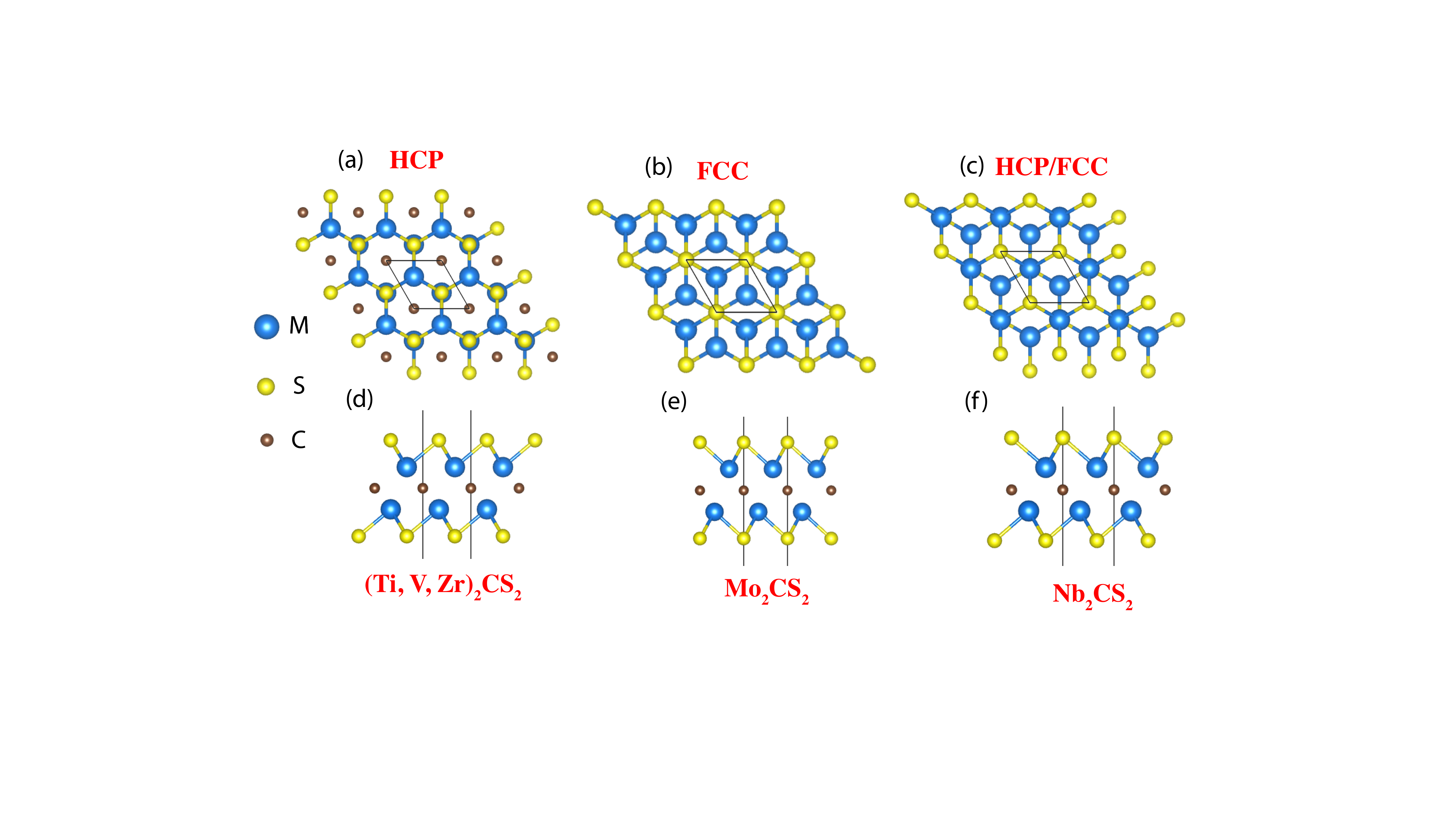}
\caption{Top view of M$_2$CS$_2$ monolayers with (a) HCP, (b) FCC, and (c) HCP-FCC surface structures. The side views are also given for HCP in (d), FCC in (e) and HCP-FCC surface structures in (f).  The thin, solid lines indicate the unit cell.}
\label{fig_structures}
\end{figure}

We first considered different possible surface structures for M$_{2}$CS$_{2}$ (M= Ti, V, Zr, Mo, Nb) monolayers.  All of the ground-state structures of bare-M$_{2}$C are hexagonal, and consist of a C layer sandwiched between two M layers.  In a previous paper by Siriwardana \emph{et al.},\cite{deniz} three types of surface structures -  namely HCP, FCC, and HCP-FCC as shown in Fig. \ref{fig_structures} - were predicted for M$_{2}$CS$_{2}$. The Ti-, V-, and Zr-based MXenes prefer an HCP surface structure where the C atoms lie at the hollow site of the hexagons formed by M and S atoms.  The Mo based MXene forms a FCC surface structure in which the C atoms reside directly beneath (above) the top (bottom) surface S atoms. The Nb$_{2}$CS$_{2}$ tends to form a mixed HCP-FCC surface structure.

\begin{figure}
\includegraphics[width=13cm]{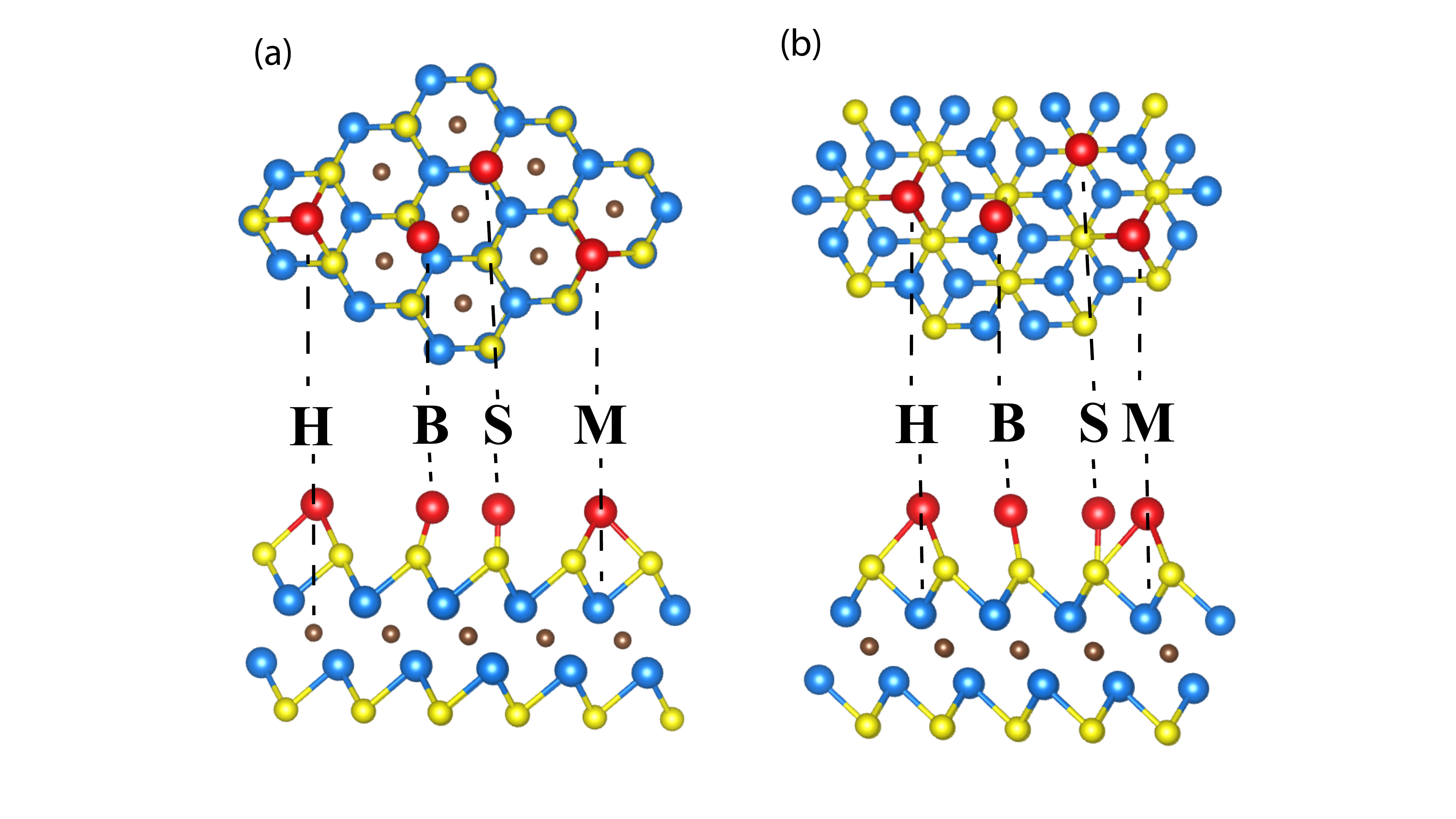}
\caption{Binding sites of (a) HCP and (b) FCC structures.  Hollow (H), Bridge (B), top of Sulfur (S), and top of transition metal (M) are featured. Red balls represent the adsorbent.}
\label{fig_sites}
\end{figure}

The adsorption energies (defined in Eq. \ref{adsEnergy}) of Mg and Na on PBE+D3 and SCAN relaxed M$_{2}$CS$_{2}$ materials are given in Tables \ref{table_primative_ads} and \ref{table_ads_4x4x1} for 1$\times$1$\times$1 (primitive) and 4$\times$4$\times$1 cells, respectively. The primitive cell represents the full-coverage case where the adatoms cover every possible binding site, while the 4$\times$4$\times$1 supercell represents a dilute case where we only have one absorbed ion and there are practically no interactions between adatoms in the periodic images.  We calculated adsorption on the hollow (H), top-metal (M), bridge (B), and top-sulfur (S) sites, as denoted in Fig \ref{fig_sites}.  
The adatoms prefer to bind most strongly to either the M or H site, depending on the MXene, adatom type, and coverage. We present  single (first row) and double side (second row) Mg and Na adsorption energies for the primitive cell in Table \ref{table_primative_ads}. Here, single (double) side means that there is a one full layer of Na/Mg absorbed on one surface (both surfaces: one layer on each surface) of the MXene structure. Since the top and bottom surface structures of Nb$_{2}$CS$_{2}$ monolayer is asymmetrical, we studied single adatom adsorption on the HCP and FCC sides separately. We found that both Mg and Na bind most strongly to the top-Nb site regardless of surface termination type. Thus, we only present the top-Nb adsorption energy for adatoms on both sides of Nb$_{2}$CS$_{2}$. For single side adsorption, from Table \ref{table_primative_ads}, it is clear that both adatoms bind least strongly to Mo$_{2}$CS$_{2}$ and most strongly to Ti$_{2}$CS$_{2}$ and Zr$_{2}$CS$_{2}$.  Most likely the adsorption properties of Mo$_{2}$CS$_{2}$ differ from that of the other examined materials because it is the only one with a purely FCC structure. Likewise, Mg and Na bind less strongly to the FCC than to the HCP side of Nb$_{2}$CS$_{2}$.  These results may be partially attributed to the lattice constants of the MXenes. Since Ti$_{2}$CS$_{2}$ and Zr$_{2}$CS$_{2}$ have some of the largest lattice constants (3.449 {\AA} for Zr$_{2}$CS$_{2}$ and 3.155 {\AA} for Ti$_{2}$CS$_{2}$) among the considered systems, the repulsive Coulomb interaction is lower in these systems than in Mo$_{2}$CS$_{2}$ (3.053 Å). Note however that the adatoms bind more strongly to V$_{2}$CS$_{2}$ than to Mo$_{2}$CS$_{2}$, although the former has the lowest lattice constant of all the examined MXenes (3.06 Å).  Also Nb$_{2}$CS$_{2}$ has a higher lattice constant (3.251 Å) than Zr$_{2}$CS$_{2}$, but a smaller adsorption energy for both sides, which is likely related to Nb$_{2}$CS$_{2}$'s hybrid structure.  Overall, we found similar patterns for the single and double sided cases.  The adsorption energies per metal atom are similar for single and double side cases, meaning that the interaction between the metal atoms (Na or Mg) on the opposite surfaces is small and is effectively screened by the MXene.

\begin{table}[!htbp]
\centering
\begin{adjustbox}{width=0.8\textwidth}
\small
\begin{tabular}{rrrrrrrrrrrrr}
  \hline
 Structure & Mo$_{2}$CS$_{2}$ & Ti$_{2}$CS$_{2}$ & V$_{2}$CS$_{2}$ & Zr$_{2}$CS$_{2}$ & Nb$_{2}$CS$_{2}$ HCP/FCC \\ 
\hline

Mg & -1.606 & -2.196 & -1.845 & -2.210 & -1.872/-1.665 \\
   & -1.608 & -2.238 & -1.904 & -2.230 & -1.790 (M) \\
   & H & M & M & H & M/M \\ \\
        
Na & -1.280 & -1.900 & -1.514 & -2.158 & -1.742/-1.546 \\
   & -1.427 & -2.088 & -1.703 & -2.229 & -1.752 (M) \\
   & M & M & M & H & M/M  \\

\hline

\end{tabular}
\end{adjustbox}

\caption{Monolayer adsorption on a M$_{2}$CS$_{2}$ primitive cell. Strongest single and double adatom adsorption energies (eV) are in the first and second row of each structure's data, and were done with PBE+D3. The corresponding lowest energy binding sites (M = top of transition metal, H = hollow site) are shown in the third row.}
\label{table_primative_ads}
\end{table} 

\begin{table}[!htbp]
\centering
\begin{adjustbox}{width=0.8\textwidth}
\small
\begin{tabular}{rrrrrrrrrrrrr}
  \hline
 Structure & Mo$_{2}$CS$_{2}$ & Ti$_{2}$CS$_{2}$ & V$_{2}$CS$_{2}$ & Zr$_{2}$CS$_{2}$ & Nb$_{2}$CS$_{2}$ HCP/FCC \\ 
\hline

Mg & -1.332 & -3.208 & -2.027 & -3.735 & -2.357/-1.609 \\
   & -1.369 & -3.378 & -2.152 & -3.847 &  -2.341/-1.566 \\
   & M & H & M & H & M/M \\ \\
   
Na & -2.520 & -2.944 & -2.778 & -3.622 & -2.938/-2.596 \\
   & -2.563 & -3.416 & -2.744 & -3.623 &  -2.928/-2.573 \\
   & M & M & M & H & M/M \\

\hline

\end{tabular}
\end{adjustbox}

\caption{Single adatom adsorption on a 4$\times$4$\times$1 supercell M$_{2}$CS$_{2}$ monolayer. Strongest PBE+D3 and SCAN adsorption energies (eV) are in first and seconds rows of each structure's data.  The corresponding binding sites (M = top of transition metal, H = hollow site) is in the third row, and are the same for PBE+D3 and SCAN calculations. }
\label{table_ads_4x4x1}
\end{table} 

We also calculated the adsorption energies for the dilute case on a 4$\times$4$\times$1 supercell structure using two exchange correlation functionals, namely PBE+D3 (first row) and SCAN (second row) as presented in Table \ref{table_ads_4x4x1}.  For most of the examined adatom-MXene combinations, the geometric relaxations with SCAN and PBE+D3 yield similar binding energies, with SCAN yielding a slightly stronger binding for all materials except for Mg/Na on either side of Nb$_{2}$CS$_{2}$ and for Na on V$_{2}$CS$_{2}$. The trends between materials are the same for both DFT functionals.  Note also that the trends for this dilute case are the same as that of the full-coverage case.  Furthermore, the relationships between adsorption, structure, and type of transition-metal are consistent with literature for Li adsorbed to the same MXenes. \cite{deniz} 

From comparison of Tables \ref{table_primative_ads} and \ref{table_ads_4x4x1}, it is evident that Mg and Na tend to bind less strongly to the MXenes in the full coverage case, than in the dilute case (except for Mg on Mo$_2$CS$_2$ monolayers).  After adsorption, Mg and Na donate charge to the MXene and thus becomes positively charged.  In these full-coverage (primitive cell) cases, the separation between these positively charged adatoms is at minimum, which enhances the repulsive Coulomb interaction.  Furthermore, in this dense limit, a reduced charge transfer from Mg and Na to MXene lowers the attractive interaction of these metal ions to the MXene. 
Finally, we found that the lowest energy binding sites are different between full-coverage and dilute cases for the Mg atoms on Mo$_2$CS$_2$ and Ti$_2$CS$_2$. For instance, Mg binds most strongly to the top-Ti site of fully-covered Ti$_{2}$CS$_{2}$, but most strongly to the hollow site of the dilute structure. We took into account this change as a function of Mg concentration when we calculated the average voltage over the adatom/MXene bilayer. 



\begin{table}[!htbp]
\centering
\begin{adjustbox}{width=0.8\textwidth}
\small
\begin{tabular}{rrrrrrrrrrrrr}
  \hline
 Structure & Mo$_{2}$CS$_{2}$ & Ti$_{2}$CS$_{2}$ & V$_{2}$CS$_{2}$ & Zr$_{2}$CS$_{2}$ & Nb$_{2}$CS$_{2}$ HCP/FCC \\ 
\hline

Mg & 0.185 & -0.404 & -0.053 & -0.418 & -0.080/0.127 &  \\
   & 0.341 & 0.797 & 0.625 & 0.694 & 0.694/0.628 &  \\
   & 2.971 & 2.619 & 2.713 & 2.636 & 2.707/2.745 &  \\
   & H & M & M & H & M/M \\ \\
           
Na & -0.061 & -0.682 & -0.296 & -0.940 & -0.523/-0.327 &  \\
   & 0.331 & 0.525 & 0.463 & 0.476 & 0.476/0.444 &  \\
   & 2.919 & 2.712 & 2.775 & 2.696 & 2.769/2.828 &  \\
   & M & M & M & H & M/M  \\

\hline

\end{tabular}
\end{adjustbox}

\caption{Primitive cell data calculated with PBE+D3.  Formation energies of M$_{2}$CS$_{2}$ - adatom structures (first row, eV), charge transfer from adatom (second row, e$^{-}$), average distance of adatom from monolayer S atoms (third row, Å), lowest energy adsorption site (fourth row).} 
\label{table_formationEng_primative}
\end{table}

\begin{table}[!htbp]
\centering
\begin{adjustbox}{width=0.8\textwidth}
\small
\begin{tabular}{rrrrrrrrrrrrr}
  \hline
 Structure & Mo$_{2}$CS$_{2}$ & Ti$_{2}$CS$_{2}$ & V$_{2}$CS$_{2}$ & Zr$_{2}$CS$_{2}$ & Nb$_{2}$CS$_{2}$ HCP/FCC \\ 
\hline

Mg & 0.455 & -1.422 & -0.240 & -1.948 & -1.393/-1.051 &  \\
   & 1.129 & 1.572 & 1.482 & 1.587 & 1.447/1.305 &  \\
   & 2.500 & 2.316 & 2.360 & 2.329 & 2.390/2.438 &  \\
   & M & H & M & H & M/M \\ \\
           
Na & 0.128 & -1.726 & -1.559 & -2.403 & -1.720/-1.378 &  \\
   & 0.862 & 0.863 & 0.862 & 0.860 & 0.860/0.863 &  \\
   & 2.712 & 2.703 & 2.697 & 2.691 & 2.711/2.715 &  \\
   & M & M & M & H & M/M &  \\

\hline

\end{tabular}
\end{adjustbox}

\caption{4$\times$4$\times$1 supercell data with PBE+D3.  Formation energies of M$_{2}$CS$_{2}$ - adatom structures (first row, eV), charge transfer from adatom (second row, e$^{-}$) average distance of adatom from monolayer S atoms (third row, Å), the lowest energy adsorption sites (fourth row).} 
\label{table_formationEng_4x4x1}
\end{table} 

Another way to test a system's stability is through the formation energy, defined in Eq. \ref{formationEnergy} in the Computational Methods section. Tables \ref{table_formationEng_primative} and \ref{table_formationEng_4x4x1} include formation energies as well as related data for full coverage (primitive cell) and dilute (4$\times$4$\times$1 supercell) cases.  The data in Table \ref{table_formationEng_primative} indicates that only Ti$_{2}$CS$_{2}$ and Zr$_{2}$CS$_{2}$ remain stable when fully covered with Mg atoms on one side.  However, every structure except for Mo$_{2}$CS$_{2}$ remains stable when fully covered with Na atoms on one side.  According to Table \ref{table_formationEng_4x4x1}, all of the dilute structures besides Mo$_{2}$CS$_{2}$ - Mg system are unlikely to decompose into M$_2$CS$_2$ + Na (Mg).  Overall, the formation energies of the full coverage cases are higher than those of the dilute cases. This is because of the close proximity of the adatoms in the former case, which increases their tendency to interact with each other via repulsive Coulomb interaction.  This argument is further supported by our charge transfer and adatom-monolayer distances, given in the second and the third row of each adatom presented it Tables \ref{table_formationEng_primative} and \ref{table_formationEng_4x4x1}.  We calculated the charge transfer using the Bader analysis\cite{bader1985atoms}, and adatom-monolayer distance by averaging the distance from the adatom to the three nearest S atoms.  The cases with less stable (positive or closer to zero) formation energies are correlated with less charge transfer from the adatoms, as well as larger adatom-monolayer distances.   

Almost all of the structures with Mg adsorbants have more charge transfer and shorter adatom-monolayer distances, but lower formation energies than those with Na adsorbants.  The trend for charge transfer is explained by Mg having a higher valency than Na, which means that Mg has more available charges to transfer than Na.  One of the reasons why the Mg-S distance is shorter than the Na-S distance is because of the atomic radius of the adatoms.  Na/Na$^+$ has a larger ionic radius than Mg/Mg$^{2+}$. Also, due to the larger charge transfer, Mg ion interacts more strongly with the MXene surface, giving rise to a shorter distance.  Interestingly, we see this trend in both dense (Table \ref{table_formationEng_primative}), and dilute cases (Table \ref{table_formationEng_4x4x1}), for all but Mg on Mo$_{2}$CS$_{2}$.

\subsection{Adatom Bilayers and Trilayers}

\begin{figure}
\includegraphics[width=14cm]{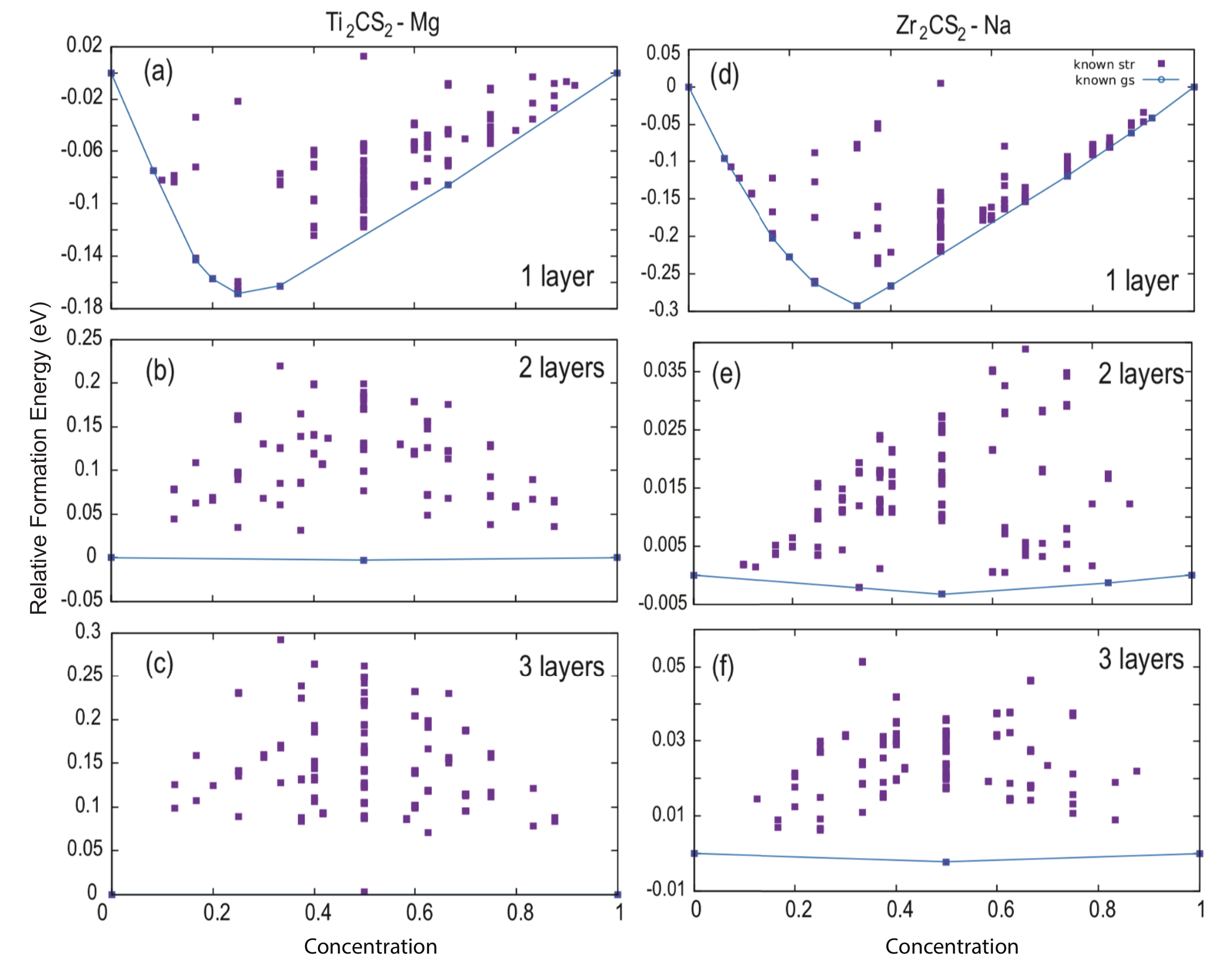}
\caption{Calculated relative formation energies as a function of concentration for up to three layers of (a-c) Mg on Ti$_2$CS$_2$ and (d-f) Na on Zr$_2$CS$_2$.  The convex hull is given by the blue curve, and lies on the lowest energy states.  The lowest energy configurations are shown in Fig. \ref{fig_CE_2}. }
\label{fig_CE}
\end{figure}

\begin{figure}
\includegraphics[width=14cm]{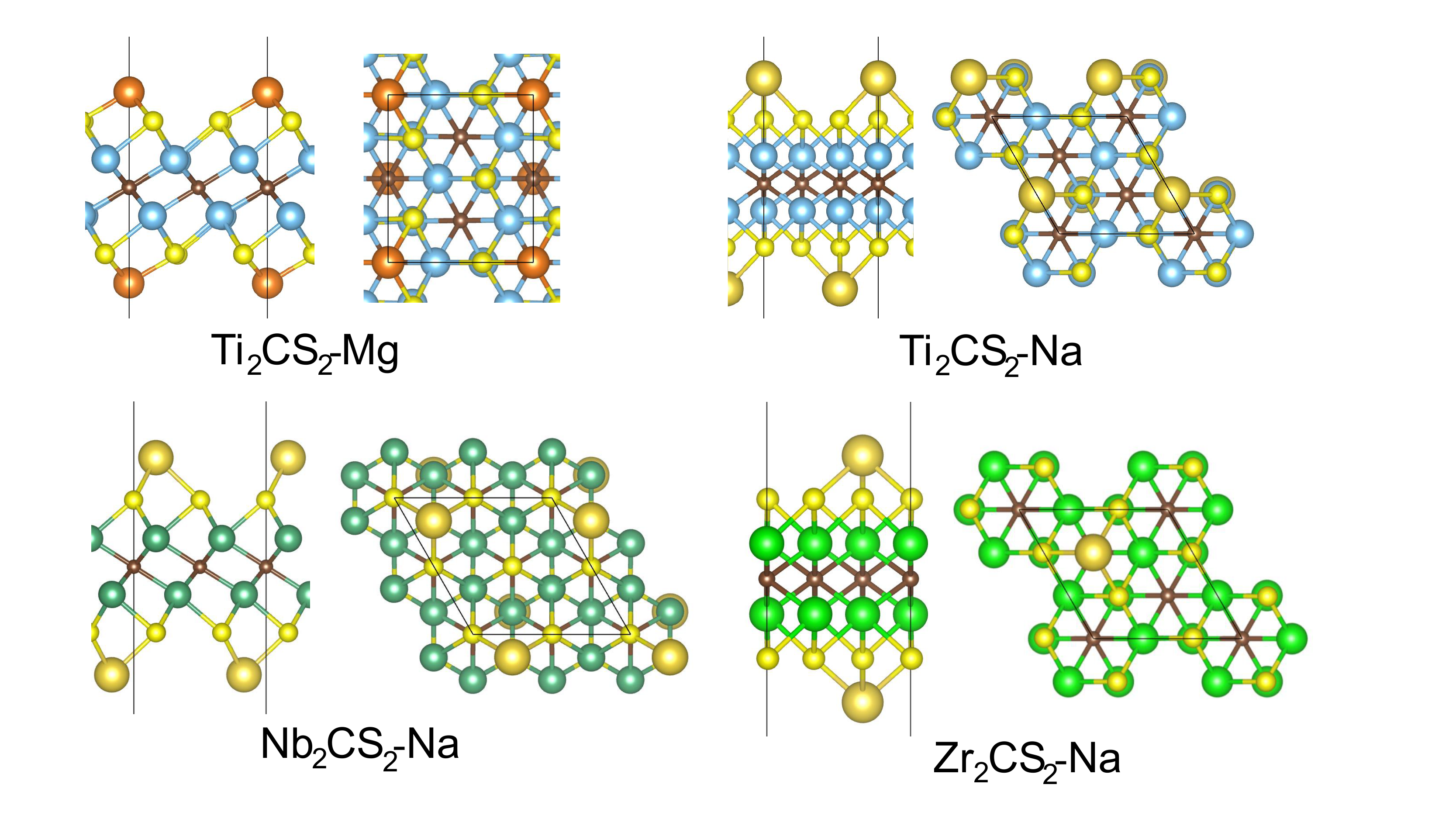}
\caption{Selected relaxed lowest formation energy structures of Fig. \ref{fig_CE}. Brown, blue, dark green, light green, yellow, dark yellow and orange balls represent C, Ti, Nb, Zr, S, Na and Mg atoms, respectively.}
\label{fig_CE_2}
\end{figure}

So far, we have only presented results for Mg and Na atoms covering at most a single layer on each surface of the considered monolayers.  To maximize the storage capacity, it is critical to prove that an electrode material can sustain multiple layers and thus provide high enough storage capacity for its battery. In this respect, we studied the energetic and voltage profiles of one, two, and three layers fully covering both surfaces of the MXenes. In the literature, the possible adsorption configurations for ions as a function of concentration have been found by checking a few different possible adsorption structures for each concentration, which may lead to incorrect voltage profiles. Also, the correct lowest energy structures for each adatom concentration may be wrongly predicted due to the existence of a limited set of possible configurations. In this study, we performed a cluster expansion\cite{CE_Zunger, CE_vandeWalle} to find the lowest energy distribution of Na/Mg atoms on MXene surfaces as a function of concentration. The energy of each considered structure was calculated using both PBE and PBE+D3 functionals, in order to assess the effect of vdw interactions on the binding mechanism. Figure \ref{fig_CE} denotes the PBE calculated relative formation energies (Eq. (8) on Ref. \cite{janus}) as a function of Na/Mg concentration obtained from the cluster expansion calculations using the ATAT code\cite{avdw:atat2,avdw:atat,avdw:maps}. In this figure, the convex hull connects the lowest energy structures that are the most likely to form in experiments. In other words, the convex hull identifies thermodynamically stable structures at T = 0 K. Figure \ref{fig_CE} includes the cluster expansion search of Ti$_2$CS$_2$-Mg and Zr$_2$CS$_2$-Na for up to three layers of Na/Mg on both surfaces of these MXenes, while Figure \ref{fig_CE_2} presents the selected lowest energy configurations (including Nb$_2$CS$_2$-Na). For these calculations, we added the second and third layers while keeping the first layers (those closest to the MXene) fixed.  The cross validation errors, measuring the predictive power of cluster expansion, are as small as 5 meV per cell, implying an accurate prediction of the convex-hull.  We considered at least 100 structures for each considered system. The relative formation energy of a single layer of Na/Mg ions on each surface of Ti$_2$CS$_2$-Mg/Zr$_2$CS$_2$-Na is more negative as compared to the two and three layer cases. As the number of layers is increased, the adsorption becomes less favorable due to the repulsive interaction between ions and increased tendency for the formation of bulk Mg and Na.  For the single layer case, the lowest energy ion concentration appears below $x$=0.4, while it is at $x$=0.5 for the two and three layers cases. We compute voltages for only the structures predicted to be on the convex hull. However, we should note that, as seen in Fig. \ref{fig_CE}, the relative formation energy of some structures appear very close to the convex hull, which can be accessible at finite temperatures.  Also, the functional may influence the shape and the number of structures on the convex-hull. Our calculations showed that the PBE and PBE+D3 calculations predicted very similar results.

\begin{figure}
\includegraphics[scale=0.4]{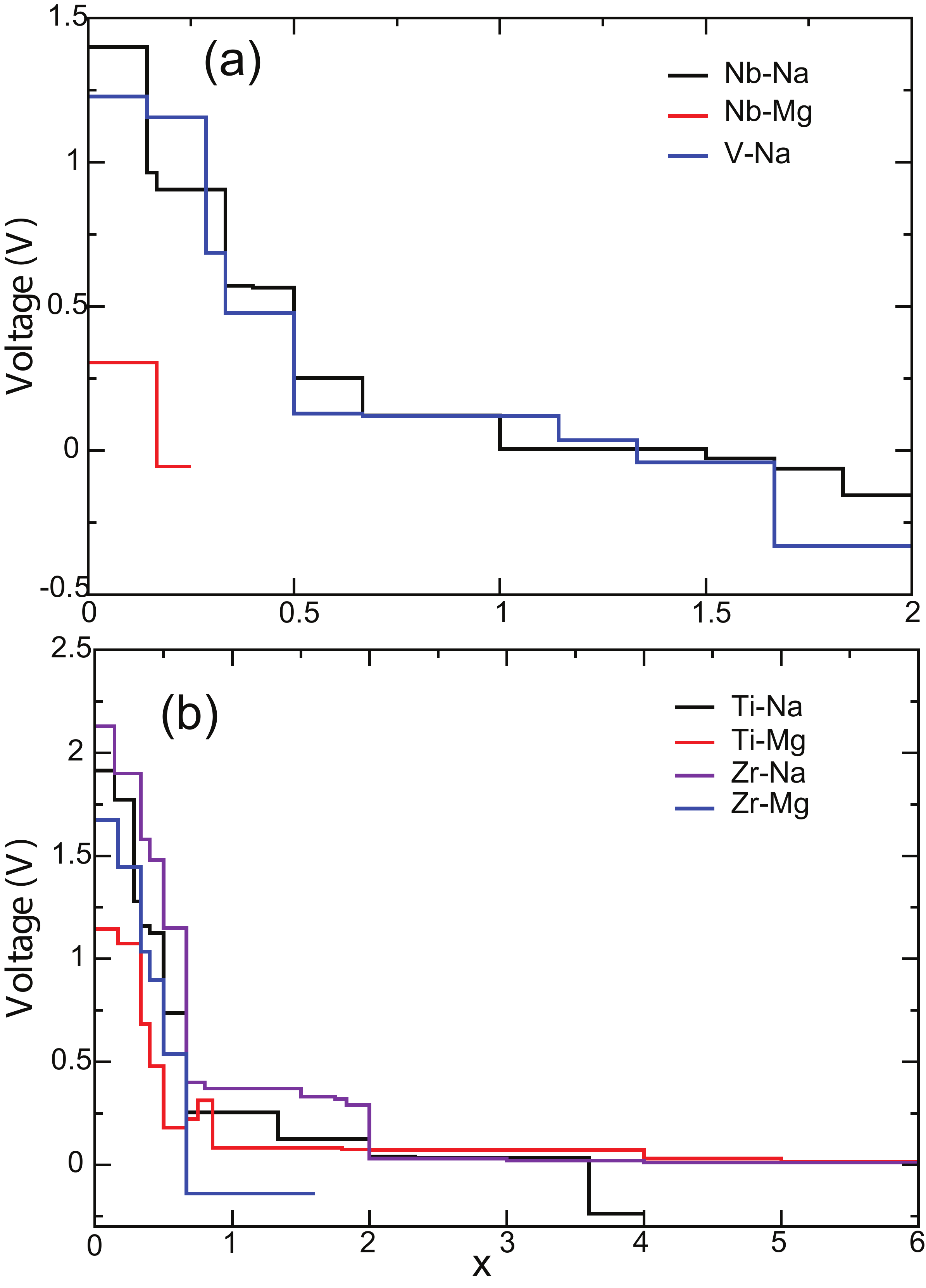}
\caption{Voltage profiles as a function of Mg and Na coverage, where $x$ is the number of adatom layers on both sides of an MXene. $x$=2 denotes one full layer of adatoms on both sides of an MXene. The results for (a) Nb$_{2}$CS$_{2}$ and V$_{2}$CS$_{2}$, and (b) Ti$_{2}$CS$_{2}$ and Zr$_{2}$CS$_{2}$ are presented. }
\label{fig_voltage}
\end{figure}

One important quantity of a battery is the cell output voltage, defined as the chemical potential difference between the cathode and anode. In this study, we considered thermodynamically stable compounds formed on the convex-hull obtained from the cluster expansion calculations.  The voltage profiles obtained from PBE calculations for the considered systems are given in Fig. \ref{fig_voltage} as a function of coverage.  We do not include results for systems with positive or close to positive formation energy, namely Mg/Na on Mo$_{2}$CS$_{2}$ and Mg on V$_{2}$CS$_{2}$.  From Fig. \ref{fig_voltage}(a), our results show that Nb$_{2}$CS$_{2}$ is inappropriate for a Mg-ion battery, since the voltage become negative after just 10$\%$ (x=0.2) of the MXene is covered by Mg on both sides.  V$_{2}$CS$_{2}$ is also not suitable as a Na-ion battery anode, as the voltage becomes negative after only 75$\%$ (x = 1.5) of the MXene is covered by Na.  However another group, Li \emph{et al}. calculated 0.49 V for one layer of Na on both sides of V$_{2}$CS$_{2}$.\cite{sulfer}  In our work, the Nb$_{2}$CS$_{2}$ performs about the same as V$_{2}$CS$_{2}$ as an Na-battery electrode when there is about 25-50$\%$ (x=0.5-1.0) of a single layer of Na atoms on both sides. There are stable phases in this range, as indicated by the voltage plateaus.  Regardless, increasing coverage beyond half of a layer drives the voltage very close to zero. After about 75$\%$ of a layer (x=1.5), the voltage becomes negative. 

The voltage profiles of Ti$_{2}$CS$_{2}$ and Zr$_{2}$CS$_{2}$ structures are shown in Fig. \ref{fig_voltage}(b).  After one full layer on both surfaces (i.e., two Na/unit cell)), the voltage of Ti$_{2}$CS$_{2}$ - Na plateaus around 0.1 V, until it drops below zero when the number of layers on each surface exceed 1.75 layers (x=3.5). Note that Wang \emph{et al.} predicted the voltage to be 0.28 V after three layers of Na on both surfaces.\cite{WANG2020227791}  This discrepancy may be due to the fact that we performed cluster expansion calculations, which enables us to determine the lowest energy adsorption structure and concentrations more accurately.  Between 0.5 and two full layers (two-four Mg/unit cell) on both surfaces, the voltage of Ti$_{2}$CS$_{2}$ - Mg plateaus at about 0.1 V, and drops only slightly after coverage is further increased.  The voltage of Zr$_{2}$CS$_{2}$ - Na decreases slightly during the formation of the first layer (x=2), but remains close to ~0.25 V.  After the first layer is completely formed on each surface, the voltage plateaus around 0.1 V, where it remains even after the third layer on each surface is fully formed. The results for Zr$_{2}$CS$_{2}$ - Mg are very different. It's voltage becomes negative even before the first layers form. Overall, these results are important, as it implies that various adatoms may behave very differently on the same electrode.  Figure \ref{fig_voltage} shows that only Nb$_{2}$CS$_{2}$ - Na/Mg, Ti$_{2}$CS$_{2}$ - Na/Mg, and Zr$_{2}$CS$_{2}$ - Na are suitable MXene-adatom combinations, because they are the only ones that can form stable multiple layers. 

Including vdW corrections does not change the achievable maximum voltage for the Mg case. However, for the Na case, we found an upward shift of the maximum voltage by about 0.1 V. For instance, for Ti$_{2}$CS$_{2}$ (Zr$_{2}$CS$_{2}$), the maximum voltage increases from 1.91 (2.13) to 2.01 (2.20) V. The concentrations for which the voltage values become negative are usually similar for both PBE and PBE+D3. For Na-covered Nb$_{2}$CS$_{2}$, PBE predicts that the voltage becomes negative when $x$ $>$ 1.8. However, Nb$_{2}$CS$_{2}$ is found to host at least one layer of Na on each surface at the PBE+D3 level.  

%
%
%

\subsection{Diffusion Across Monolayers}

\begin{figure}
\includegraphics[width=10cm]{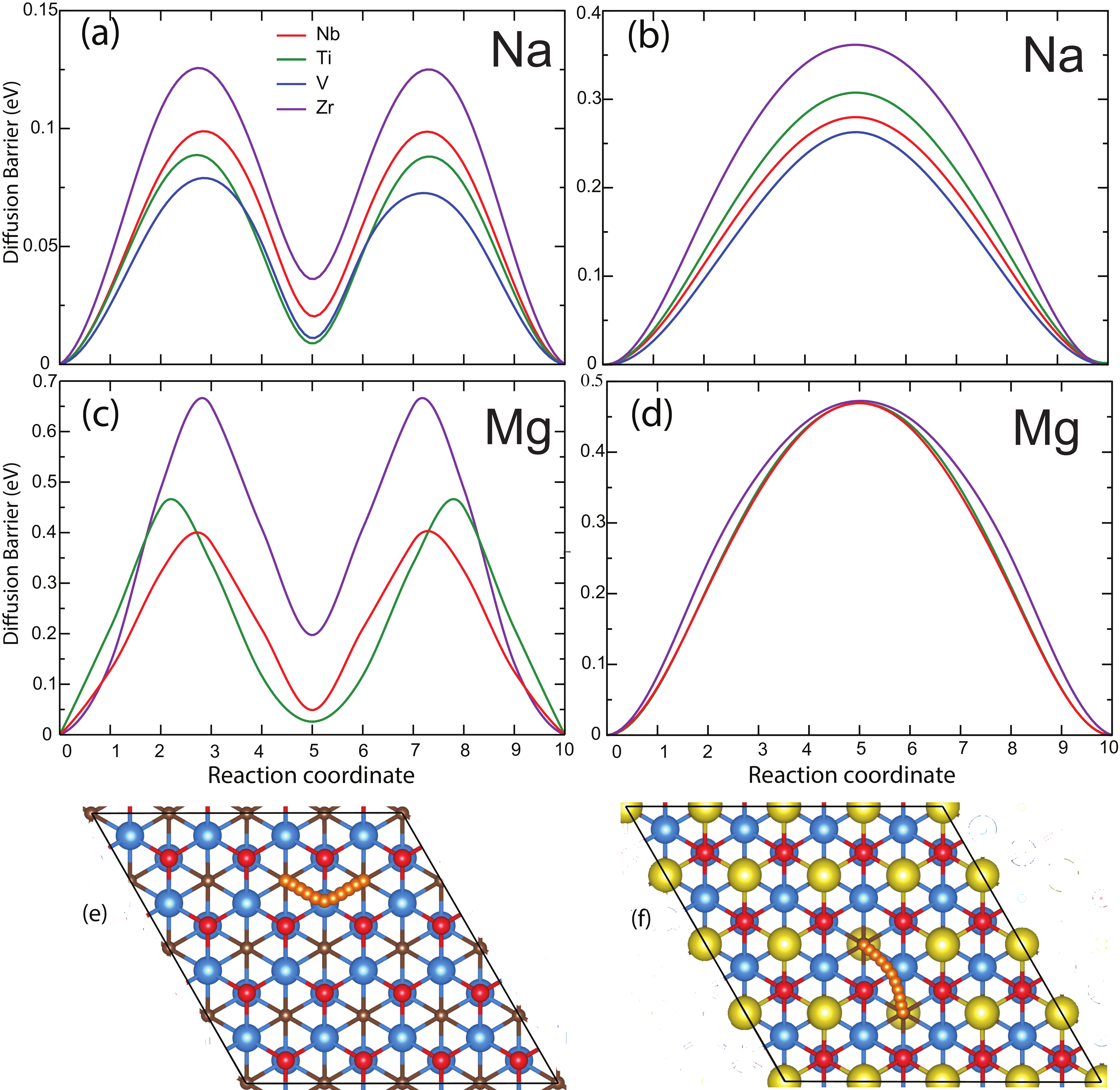}
\caption{Diffusion of Na (in (a) and (b)) and Mg (in (c) and (d)) across monolayers on a 4$\times$4$\times$1 supercell. Two cases were considered. In (a) and (c), a single adatom migrates between the lowest energy binding sites.  In the second case shown in (b) and (d), there is already a layer of adatoms present on both sides, albeit for a single point defect. Diffusion paths are given for dilute and dense coverage are given in (e) and (f), respectively. Blue, brown, red and yellow balls represent M, C, S and Na/Mg atoms, respectively. }
\label{fig_diffusion}
\end{figure}

In the operation of a battery, it is critical to understand migration of metal atoms across these candidate electrode materials. One way to do this is by calculating diffusion energy barriers that reflect how difficult it is for a particular adatom to travel across a material, specifically along the lowest energy path. We calculated diffusion energy barriers and diffusion paths using PBE and PBE+D3 functionals to uncover the influence of the vdW correction. In Fig. \ref{fig_diffusion} and Table \ref{table_diffusion}, we present the diffusion energy barriers of Mg and Na for two situations. Figure \ref{fig_diffusion} presents the results obtained from PBE. In the first case (shown in Fig. \ref{fig_diffusion}(a) and (c)), we considered a dilute system where a single metal atom migrates between the lowest energy adsorption sites on a 4$\times$4$\times$1 supercell. As noted in the adsorption energy section, the lowest energy adsorption sites are the hollow or top-M sites depending on the MXene-adatom combination. For example, Na travels across Ti$_{2}$CS$_{2}$ from top-Ti to top-Ti site (endpoints of barrier plot in Fig. \ref{fig_diffusion}(e)) while passing through a meta-stable hollow site. In contrast, the Mg atom travels across Ti$_{2}$CS$_{2}$ from hollow to hollow site while passing through a meta-stable top-Ti site. The calculated energy barriers are in the range of 0.075-0.125 eV (0.4-0.65 eV) for Na (Mg) with the PBE functional.  Mo$_{2}$CS$_{2}$ - Na and V$_{2}$CS$_{2}$ - Mg cases are excluded from this analysis as they have positive or very low formation energies, which means that adsorption is not favorable even in the dilute cases. In Table \ref{table_diffusion}, we also present the barrier values when the vdW correction is included. The barrier values are about 0.02-0.05 eV higher for PBE+D3 due to stronger interaction between the MXene surfaces and metal atoms as a result of including the vdW interaction. 

Figure \ref{fig_diffusion}(a) features diffusion energy barriers of Na for dilute doping concentrations.  Only the HCP side of Nb$_{2}$CS$_{2}$ was considered.  Na atop of V$_{2}$CS$_{2}$ and Zr$_{2}$CS$_{2}$ experiences the lowest (0.078 eV) and highest (0.124 eV) energy barriers, respectively.  This makes sense, as Na binds weakest to V$_{2}$CS$_{2}$ and strongest to Zr$_{2}$CS$_{2}$. Likewise, Figure \ref{fig_diffusion}(c) indicates that Mg has a higher diffusion barrier on Zr$_{2}$CS$_{2}$ (0.620 eV) than on the other examined materials.  Note that the energy barriers for Na are much lower than that of Mg; thus, Na diffuses much more easily on MXenes than Mg does. This is due to the fact that Mg has a higher valency than Na, so that the former interacts more strongly with the surrounding atoms and thus encounters more resistance.  Additionally, Na has a larger atomic radius and rests farther away from the monolayers than Mg, as we already mentioned in the previous section.  Our results match those of the literature well.  We predicted an almost identical energy barrier of Mg and Na on Ti$_{2}$CS$_{2}$ as Wang \emph{et al.} (0.46 eV and 0.11 eV)\cite{WANG2020227791}.  We also predicted a similar energy barrier of Na on V$_{2}$CS$_{2}$ as Li \emph{et al.} (0.06 eV)\cite{sulfer}. When one compares Li atom diffusion barriers on these MXenes with those for Mg and Na,  it appears that for every MXene Mg experiences a higher and Na a lower energy barrier than Li\cite{deniz}.

In Fig. \ref{fig_diffusion} (b) and (d), we have a 4$\times$4$\times$1 supercell of considered MXenes on which all of the lowest energy adsorption sites, except one (a vacant site), are occupied by  either Na or Mg atoms on both surfaces. This corresponds the second situation or dense case. We present our results for a metal atom migrating into this vacancy from a nearby occupied adsorption site. As shown in Fig. \ref{fig_diffusion}(f), the metal atom hops over a crossing site (a mid point between bridge and second most lowest energy adsorption site) on its way to the vacancy. The ordering of the energy barriers are consistent with adsorption energies of full coverage cases, shown in Table \ref{table_primative_ads}. Note that most of the energy barriers for the full-coverage cases are higher than those for dilute coverage. This is because the metal atom is forced to travel near the bridge site, an energetically unstable point and because of the repulsive coulomb interactions from the nearby positively charged metal ions. Overall, it appears that Na travels easiest on V$_{2}$CS$_{2}$ (with energy barrier of 0.263 (0.282) eV with PBE (PBE+D3)), and Mg on HCP-Nb$_{2}$CS$_{2}$. Mg on Zr$_{2}$CS$_{2}$ has a relatively high diffusion energy barrier.  Comparing the Li atom diffusion on the considered MXenes for a dense case, it appears that, for every MXene, Mg experiences a higher and Na a lower energy barrier than Li\cite{deniz}.

\begin{table}[!htbp]
\centering
\begin{tabular}{rrrrrrrrrrrrr}
  \hline
 Metal ion & Ti$_{2}$CS$_{2}$~~~~~ & Zr$_{2}$CS$_{2}$~~~~~ & Nb$_{2}$CS$_{2}$~~~~~ \\ 
\hline
Mg~~~~ & 0.450 (0.468)                & 0.620 (0.645)    & 0.382 (0.403)  \\
   & 0.469 (0.520)                & 0.473 (0.514)    & 0.469 (0.511)  \\ 
Na~~~~ & 0.087 (0.095)                & 0.124 (0.136)    & 0.098 (0.106)   \\
   & 0.307 (0.325)                & 0.362 (0.381)    & 0.280 (0.305)  \\



\hline

\end{tabular}

\caption{Diffusion barriers are given in eV. For each metal atom (Mg and Na),  the first number is for the dilute case and the second number is for the dense case.  Only the result for HCP side of Nb$_{2}$CS$_{2}$ is provided. The numbers in parenthesis are the barrier values when including the vdw correction within DFT + D3 scheme. } 
\label{table_diffusion}
\end{table}

%
%
%

\subsection{Molecular Dynamics}

\begin{figure}
\includegraphics[width=16cm]{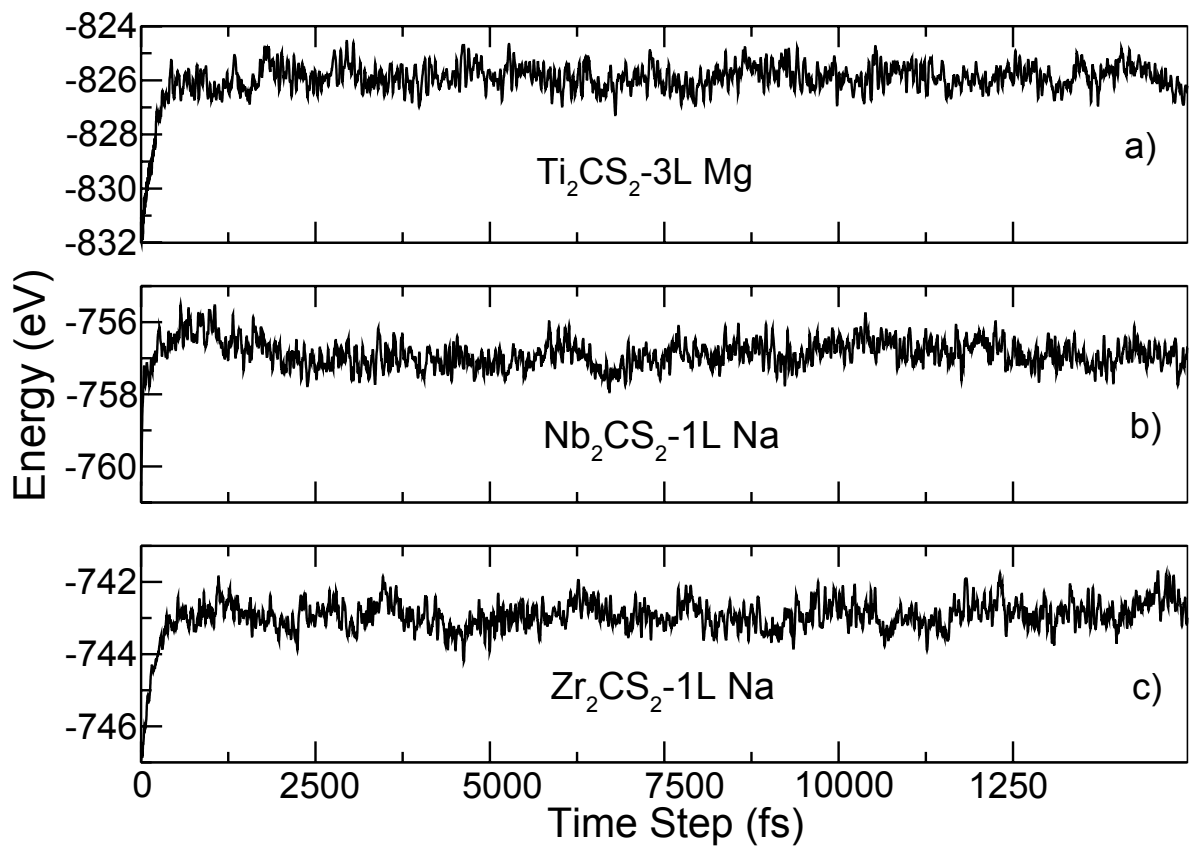}
\caption{Energy vs time step of molecular dynamics simulations at T=300 K for (a) three layers of Mg on Ti$_{2}$CS$_{2}$, (b) one layer of Na on Nb$_{2}$CS$_{2}$,  and (c) one layer of  Na on Zr$_{2}$CS$_{2}$.  }
\label{fig_MD_timestep}
\end{figure}

\begin{figure}
\includegraphics[width=16cm]{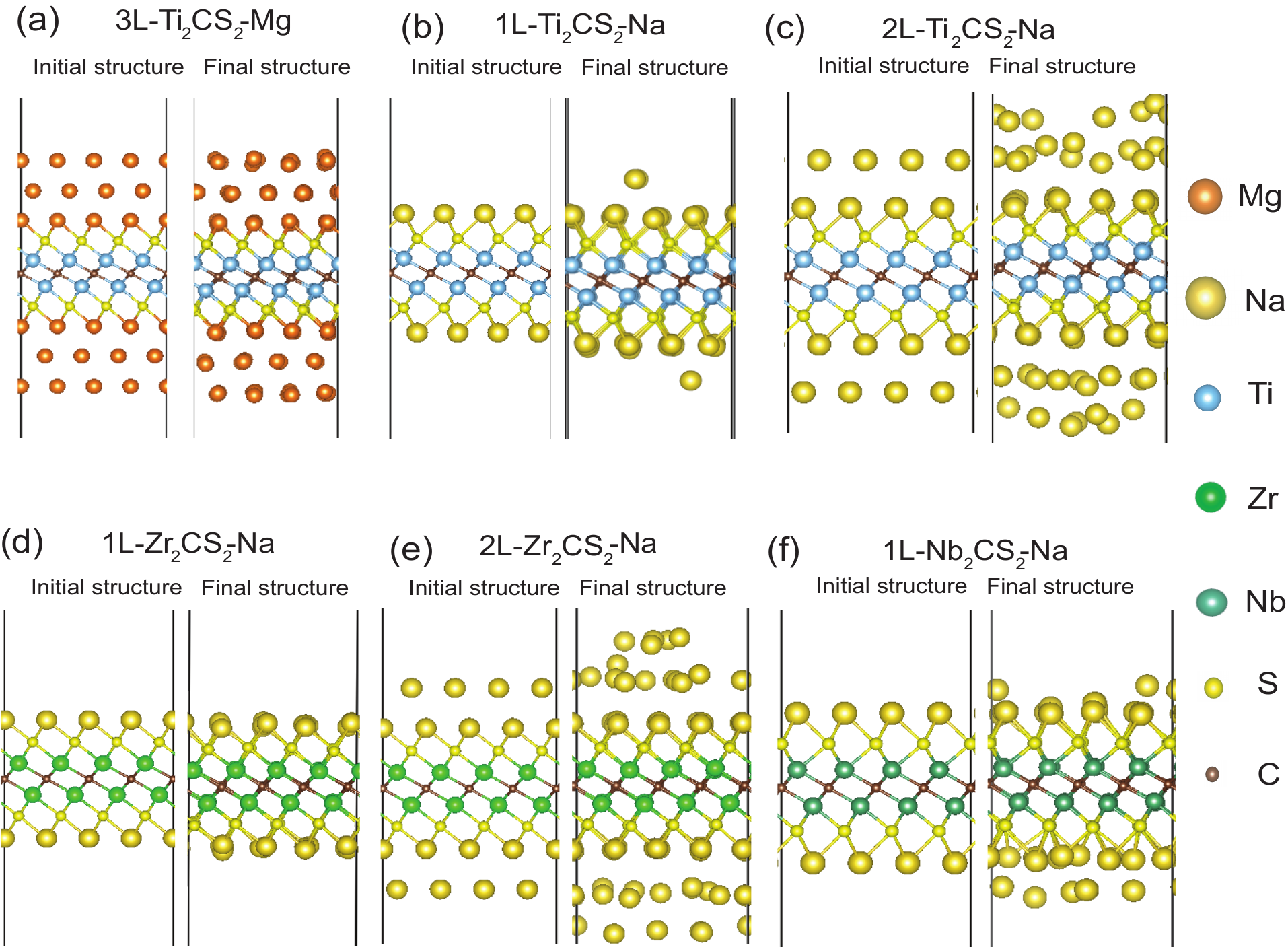}
\caption{Molecular dynamics simulations at T=300 K for (a) three layers of Mg on Ti$_{2}$CS$_{2}$, (b) one layer of  Na on Ti$_{2}$CS$_{2}$, (c) two layers of Na on Ti$_{2}$CS$_{2}$, (d) one layer of Na on Zr$_{2}$CS$_{2}$, (e) two layers of  Na on Zr$_{2}$CS$_{2}$ and (e) one layer of Na on Nb$_{2}$CS$_{2}$. For each case, initial and final structure after a 15 ps simulation are given. }
\label{fig_MD}
\end{figure}

We ran molecular dynamics (MD) simulations lasting 15 ps with a time step of 1 fs at T= 300 K for Mg and Na on Ti$_{2}$CS$_{2}$ and for Na on Zr$_{2}$CS$_{2}$ and Nb$_{2}$CS$_{2}$. Figure \ref{fig_MD_timestep} shows the variation of energy as a function of time step in the selected systems. The energy oscillates around an average value with a magnitude of $\sim$20 meV/Mg or Na. We also determined what the PBE and PBE+D3 functionals predict for the stability of the considered systems at room temperature. According to our voltage calculations done at 0 K with PBE and PBE+D3, Ti$_{2}$CS$_{2}$ and Zr$_{2}$CS$_{2}$ remain stable after multiple layers of Na and Mg adsorption.  We considered one, two and three layers of ions on both surfaces of 4$\times$4$\times$1 supercell structures in order to determine stability of the respective systems against thermal effects. The initial and  final structures in our MD simulations performed with PBE+D3 are shown in Fig \ref{fig_MD}. MD calculations indicate that three layers of Mg on both surfaces of Ti$_{2}$CS$_{2}$ is stable for both PBE and PBE+D3 functionals, meaning that Mg layers remain intact on Ti$_{2}$CS$_{2}$, thereby offering high capacity, stable Mg storage for up to three layers. In addition, we did not observe any significant structural changes for the Ti$_{2}$CS$_{2}$ monolayer. In contrast, for a single layer of Na on both surfaces of Ti$_{2}$CS$_{2}$,  
some of the Na atoms leave the surface and form a second layer in our 15 ps simulation at 300 K. In spite of the voltage calculations, the full coverage of Ti$_{2}$CS$_{2}$ surfaces with Na is not stable at room temperature. For the two-layer case in which two layers of Na are placed on both surfaces of Ti$_{2}$CS$_{2}$, the Na atoms in the second layer wiggle a lot and eventually detach from the MXene surface. This is in contradiction with the study of Wang \emph{et al.}\cite{WANG2020227791} who claimed from voltage data only that Ti$_{2}$CS$_{2}$ could support three layers of Na on both surfaces. Undesired release of ions from the electrode without a connection to the external circuit certainly leads to fast decay of capacity. Our MD simulations highlight the importance of using multiple methods to verify a structure's stability. 

At the PBE level, despite the formation energy and voltage values implying the stability of the considered systems, two and three layers of Na ions on Zr$_{2}$CS$_{2}$ exhibit detachment of the outer most Na layers due to weak binding of these layers with the MXene surface. This is also evident from the voltage results where we found that the value of voltage becomes very low when Zr$_{2}$CS$_{2}$ starts to accept multi-layers on each surface. We also observed a partial release of Na ions for the one-layer case from  the Zr$_{2}$CS$_{2}$ surface. However, such a partial detachment of Na ions obtained from PBE calculations disappears in the case of the PBE+D3 functional. While both PBE and PBE+D3 functionals predict the similar results (such as the maximum number of layers absorbed on the each surface and voltage values), the inclusion of a vdW correction is necessary to predict finite temperature stability correctly. Our PBE and PBE+D3 calculated voltage values point out that three-layer Na  absorption is possible for Zr$_{2}$CS$_{2}$. However, MD simulations computed with PBE and PBE+D3 indicates a complete detachment of second and third layers of Na.  Finally, we simulated Nb$_{2}$CS$_{2}$-Na system. Similar to the Ti$_{2}$CS$_{2}$-Na case, Na atoms detach from the Nb$_{2}$CS$_{2}$ surfaces. 

Finally, to further understand the effect of temperature on the binding energies, we considered the initial and final structures of MD simulations for Ti$_2$CS$_2$-3L Mg (Fig. \ref{fig_MD} (a) ), Zr$_2$CS$_2$-1L Na (Fig. \ref{fig_MD} (d)) and  Nb$_2$CS$_2$-1L Na (Fig. \ref{fig_MD} (f)). When we compare energy of initial and final structures, the binding energies (eV/Mg or eV/Na) may change up to 164 meV. Specifically, the changes are 61 meV/Mg for Ti$_2$CS$_2$-3L Mg, 145 eV/Na for Zr$_2$CS$_2$-1L Na and  164 meV/Na for Nb$_2$CS$_2$-1L Na.

\section{Conclusion}
We assessed the potential of M$_{2}$CS$_{2}$ monolayers (where M= Mo, Ta, V, Zr) as electrode materials for Na and Mg batteries. We found that Mo and V-based MXenes are not suitable as anode materials due to their limited storage capacities. However, Ti and Zr-based MXenes promise high-capacity for Mg and Na batteries. Ti$_{2}$CS$_{2}$ can accommodate three layers of Mg on its surfaces. Similarly, the adsorption of three layers of Na ions is energetically possible for both surfaces of Zr$_{2}$CS$_{2}$. In the case of Na-Ti$_{2}$CS$_{2}$, one and half layers of Na on both surfaces can stay energetically stable. Mg ions have 0.2-0.3 eV higher diffusion barriers than Na ions. Among the considered systems, the Zr-based MXene has the highest barrier energies due to the strong binding of Na and Mg ions on the surfaces of this MXene. Our molecular dynamics simulations, lasting at least 15 ps, revealed that three layers of Mg on both surfaces of Ti$_{2}$CS$_{2}$ remain intact at 300 K with some deviation of atoms from their 0 K positions. However, Na layers are released from the Ti$_{2}$CS$_{2}$ monolayer, suggesting that encapsulation may be needed to preserve Na layers. In spite of energetic and voltage calculations, Zr$_{2}$CS$_{2}$ can only support one layer of Na at 300 K. This is because, according to our MD simulations, the multilayer adsorption of Na is unlikely on the Zr$_{2}$CS$_{2}$ surfaces.  Our calculations highlight that not only binding energy and voltage calculations but also molecular dynamics simulations are essential to determine the storage capacity and stability of the corresponding systems.  

\section{Acknowledgment}
Computational resources were partially provided by HPC infrastructure of the University of Antwerp (CalcUA) a division of the Flemish Supercomputer Center (VSC), which is funded by the Hercules foundation.  The other calculations have been carried out at UMBC High Performance Computing Facility (HPCF).  This work was supported by the National Science Foundation through Division of Materials Research under NSF DMR-1726213 Grant. 
\bibliography{references}
\end{document}